\documentclass[a4paper,UKenglish,cleveref, autoref, thm-restate]{lipics-v2021}

\title{PACE Solver Description: twin\_width\_fmi} 

\titlerunning{PACE Solver Description: twin\_width\_fmi} 

\author{David Balaban}{International Computer High School of Bucharest, Romania}{david.c.balaban@gmail.com}{}{}
\author{Adrian Micl\u au\c s}{Faculty of Mathematics and Computer Science, University of Bucharest, Romania}{adrian.miclaus@fmi.unibuc.ro}{}{}

\authorrunning{David Balaban et al.} 

\Copyright{David Balaban, Adrian Micl\u au\c s} 

\ccsdesc[100]{Theory of computation~Graph algorithms analysis} 

\keywords{dominating-set, heuristics, adaptive-algorithm, greedy, local-search} 

\category{} 

\relatedversion{} 

\supplement{The source code can be shared upon request \\
GitHub (https://github.com/DavidBalaban01/Pace-2025)}

\nolinenumbers 

\EventLongTitle{10th Parameterized Algorithms and Computational Experiments(PACE 2025)}
\EventShortTitle{PACE 2025}
\EventAcronym{PACE}
\EventYear{2025}
\ArticleNo{ }
\usepackage{amsmath}
\usepackage{amsfonts} 
\usepackage{mathtools}
\usepackage{algorithm}
\usepackage{algpseudocode}
\begin{document}

\maketitle

\begin{abstract}
In this paper we present \texttt{twin\_width\_fmi}'s solver for the heuristic track of PACE's 2025 competition on Minimum Dominating Set. 

As a baseline, we implement \texttt{greedy-ln}, a standard greedy dominating-set heuristic that repeatedly selects the vertex that newly dominates the largest number of currently undominated vertices. We then use this greedy solution as the starting point for a simulated annealing local search: we attempt vertex removals and exchanges and accept worsening moves with decaying probability, in order to escape local minima while preserving domination.

Our best-performing component, which we ultimately submitted, is \texttt{hedom5}. The design of \texttt{hedom5} is inspired by recent iterative-greedy style domination heuristics~\cite{IterativeGreedy22} that alternate between constructive steps, pruning, and focused repair rather than relying on a single pass. In \texttt{hedom5}, the input graph is first stored in a compact CSR structure and simplified using fast reductions such as forcing neighbors of leaves and handling isolates. We then run a lazy gain-based greedy stage using a priority queue: each candidate vertex is scored by how many currently undominated vertices its closed neighborhood would newly dominate, and scores are only recomputed when necessary. After this constructive phase, we perform an aggressive backward pruning pass that iterates over the chosen dominators in reverse insertion order and deletes any vertex whose closed neighborhood is still fully dominated by the remaining set. Finally, we run a budgeted 1-swap local improvement step that attempts to replace a dominator by an alternative vertex that covers all of its uniquely covered vertices, thereby reducing the size of the dominating set. A brief safety patch at the end guarantees full domination.

We compare these methods on the official PACE 2025 benchmark instances and observe that while greedy+\ simulated annealing improves over the plain greedy baseline, \texttt{hedom5} consistently achieves the smallest dominating sets within the competition constraints, and was therefore selected as our submitted solver.

\end{abstract}


\section{Introduction}

The \emph{Minimum Dominating Set} (MDS) problem is a classical graph optimization problem. 
Given a graph $G = (V,E)$, the task is to find a set $D \subseteq V$ of minimum size such that every vertex in $V$ is either in $D$ or has a neighbor in $D$. 
Intuitively, one may view $D$ as a minimum set of ``stations'' that collectively monitor or serve the entire network in one hop. 
This captures a wide range of applications, for example: placing the smallest number of sensors so that every device in a communication network is observed; selecting a minimum number of access points in wireless coverage; or choosing a minimal set of representatives that can directly reach everyone else in a social/interaction graph. 
Because such coverage and control tasks arise in practical settings ranging from infrastructure planning to surveillance to information spread, high-quality heuristics for MDS can directly translate into resource savings.

From a theoretical point of view, MDS is well known to be NP-hard \cite{GareyJohnson79}, and, unless $\mathrm{P}=\mathrm{NP}$, it admits no polynomial-time $(1-\alpha)\ln n$-approximation for any constant $\alpha>0$ \cite{Chvatal79,Feige98}, which essentially matches the logarithmic-factor guarantee given by the standard greedy set-cover-style heuristic \cite{Chvatal79,Johnson74}. 
In the parameterized setting, deciding whether a graph has a dominating set of size at most $k$ is W[2]-hard \cite{DowneyFellows99}, making fixed-parameter tractability unlikely on general graphs. 
On the positive side, there is extensive structure-specific work: for planar graphs and other sparse minor-closed classes, Polynomial-Time Approximation Schemes (PTASes) and subexponential algorithms are known \cite{Baker94,FominThilikos06}, and kernelization results give provably size-reduced equivalent instances under certain restrictions \cite{AlberNiedermeier00}. 
This long line of research has made MDS a central benchmark in exact, parameterized, and approximation algorithms.

Despite this progress, solving large real-world instances under strict time limits remains difficult. 
Exact methods such as integer programming or branch-and-reduce can produce optimal dominating sets, but often become impractical at scale. 
Conversely, purely greedy algorithms are extremely fast but can get stuck in locally dense regions of the graph, yielding unnecessarily large dominating sets. 
This motivates the design of \emph{hybrid heuristics}: algorithms that combine fast constructive steps, local improvement, problem-specific reduction rules, and lightweight metaheuristics.

In this work we describe our solver, which mixes classic greedy construction, simulated annealing starting from that greedy solution, and an aggressive reduction--build--prune--swap pipeline inspired by recent iterative-greedy methods that repeatedly repair and shrink a candidate solution rather than committing to a single pass \cite{IterativeGreedy22}. We show experimentally that this combination consistently outperforms plain greedy and standalone local search on the PACE 2025 instances, and in particular that our \texttt{hedom5} component achieves our best dominating sets.

\section{Solver Description: \texttt{hedom5}}

Our main submitted solver, \texttt{hedom5}, is a multi-stage heuristic designed to construct a small dominating set quickly, aggressively simplify it, and then locally improve it under a time/attempt budget. The solver is built for the DIMACS-like \texttt{.ds} format used in PACE (1-indexed vertices), and is engineered to run in a single pass over the input plus a handful of linear-time improvement phases.

At a high level, \texttt{hedom5} proceeds in four conceptual phases:
\begin{enumerate}
    \item \textbf{Graph parsing and representation.}
    \item \textbf{Greedy construction with reductions.}
    \item \textbf{Pruning (redundancy removal).}
    \item \textbf{Local improvement via budgeted swaps.}
\end{enumerate}
A final safety patch ensures the solution is a valid dominating set.

\subsection{Graph Representation}
We store the graph in a compact Compressed Sparse Row (CSR) structure. For each vertex $v$, we maintain:
\begin{itemize}
    \item its degree,
    \item an offset \texttt{off[v]} into a flat adjacency array \texttt{nbr}, and
    \item \texttt{nbr[off[v]..off[v+1])}, which lists its neighbors.
\end{itemize}
This layout supports fast iteration over the open neighborhood $N(v)$ and the closed neighborhood $N[v] = \{v\} \cup N(v)$, which is critical because dominating-set reasoning is almost entirely neighborhood-based. Self-loops are ignored during construction.

\subsection{Stage 0: Reductions}
Before any greedy selection, we apply two cheap graph-theoretic reduction rules:
\begin{itemize}
    \item \textbf{Isolates.} If a vertex $v$ has degree $0$, then $v$ must be in every dominating set, so we add $v$ immediately.
    \item \textbf{Leaf rule.} If $u$ is a leaf (degree $1$) with unique neighbor $v$, and $u$ is still undominated, then $v$ is forced into the dominating set. Intuitively, the only way to dominate $u$ is to pick $v$, so we add $v$.
\end{itemize}
We maintain two Boolean arrays:
\begin{itemize}
    \item \texttt{dominated[v]}: whether $v$ is currently dominated,
    \item \texttt{inD[v]}: whether $v$ is currently in the dominating set $D$.
\end{itemize}
Whenever we add a vertex $v$ to $D$, we immediately mark $v$ and all its neighbors as dominated and keep track of how many vertices remain undominated.

\subsection{Stage 1: Lazy Greedy Construction}
After reductions, we build an initial dominating set using a priority-queue greedy strategy. Each candidate vertex $v$ is given a score equal to the number of currently \emph{undominated} vertices in its closed neighborhood $N[v]$. Intuitively, we want to pick the vertex that ``covers the most uncovered mass.''

Instead of recomputing exact gains from scratch after every insertion, we use a \emph{lazy} scheme:
\begin{enumerate}
    \item Initialize a max-priority queue with tuples $(\text{gain}(v), v)$ using a quick upper bound such as $\deg(v)+1$.
    \item Repeatedly extract the top candidate $v$.
    \item Recompute its \emph{true} current gain (how many currently undominated vertices $v$ would newly dominate).
    \item If the recomputed gain is lower than the stored key, push it back with the corrected gain.
    \item Otherwise, accept $v$: insert $v$ into $D$ and mark its closed neighborhood as dominated.
\end{enumerate}
This ``lazy refresh'' avoids expensive global rescoring and closely follows standard lazy greedy for coverage problems.

If the queue ever becomes empty but some vertices remain undominated (which is rare but theoretically possible due to stale keys), we simply pick an arbitrary undominated vertex and add it.

\subsection{Stage 2: Backward Pruning}
The greedy construction tends to over-select vertices. To shrink the set, we apply a deterministic pruning phase:
\begin{enumerate}
    \item For every vertex $x$ in the graph, we count how many chosen dominators in $D$ cover $x$ (including $x$ itself).
    \item We then walk over the vertices of $D$ in \emph{reverse insertion order}. For a candidate $v \in D$, we check whether every vertex in $N[v]$ is still covered at least twice (i.e., would remain dominated even if $v$ were removed).
    \item If so, $v$ is redundant: we remove $v$ from $D$ and decrement coverage counts accordingly.
\end{enumerate}
This pass quickly removes ``late'' vertices that were added mainly to fix small uncovered leftovers but that became unnecessary after subsequent insertions.

\subsection{Stage 3: Budgeted 1-Swap Local Search}
After pruning, we try to apply small, local improvements. The key idea is a limited \emph{1-swap} move:
\begin{itemize}
    \item Suppose $w \in D$. Some vertices of the graph might be \emph{uniquely covered} by $w$, meaning $w$ is the only dominator that covers them.
    \item We attempt to replace $w$ with another vertex $t \notin D$ that simultaneously covers all those uniquely covered vertices.
\end{itemize}
If such a $t$ exists, we can perform the swap $D \leftarrow (D \setminus \{w\}) \cup \{t\}$ without breaking domination. If this reduces the size of $D$ (for example, if $t$ can also allow further pruning), we keep the change.

We do not search the entire graph for $t$; we only consider a small candidate pool such as the closed neighborhood of $w$. We repeat this search for a bounded number of attempts or until a time budget expires. This is a classic local-improvement idea: try cheap single-vertex replacements that strictly improve the solution.

\subsection{Final Safety Patch}
As a final step, we verify that every vertex is dominated by the constructed set $D$. In the unlikely event that something became uncovered due to swaps or pruning, we greedily add a small number of extra vertices to cover any remaining gaps. This ensures that the solver always outputs a valid dominating set.

\subsection{Pseudocode}
Below we summarize the main flow of \texttt{hedom5} in simplified pseudocode.  
Notation: $N[v]$ is the closed neighborhood of $v$.

\begin{verbatim}
function HEDOM5_SOLVER(Graph G):
    build CSR structure from G
    init D := empty
    init dominated[v] := false for all v
    init inD[v] := false for all v

    // -------- Stage 0: Reductions --------
    // Isolates
    for each v:
        if degree(v) == 0:
            ADD_TO_D(v)

    // Leaf rule
    Q := all vertices u with degree(u) == 1
    while Q not empty:
        u := pop(Q)
        if not dominated[u]:
            v := unique_neighbor(u)
            if v != null:
                ADD_TO_D(v)

    // -------- Stage 1: Lazy Greedy --------
    PQ := priority queue of (score_estimate(v), v)
          where score_estimate(v) ~ degree(v)+1

    while exists undominated vertex:
        if PQ empty:
            pick any undominated vertex w
            ADD_TO_D(w)
            continue
        (score,v) := PQ.pop_max()
        if dominated[v]:
            continue
        true_gain := count_undominated_in_closed_neighborhood(v)
        if true_gain < score:
            PQ.push( (true_gain,v) )
            continue
        ADD_TO_D(v)

    // -------- Stage 2: Pruning --------
    cover_count[x] := number of vertices in D that dominate x
    for each v in D in reverse insertion order:
        if for all x in N[v]: cover_count[x] >= 2:
            remove v from D
            decrement cover_count[x] for all x in N[v]

    // -------- Stage 3: 1-Swap Local Search --------
    attempts := 0
    while attempts < ATTEMPT_CAP and time < TIME_BUDGET:
        for each w in D (heuristic order):
            U := { x : x in N[w] and w is unique dominator of x }
            if U is empty: continue
            C := candidate vertices t in N[w] \ D
            for t in C:
                if U subset of N[t]:
                    // perform swap
                    remove w from D
                    add t to D
                    recompute cover_count, unique ownership, etc.
                    break
        attempts := attempts + 1

    // -------- Final Safety Patch --------
    if some vertex x is not dominated by D:
        greedily add a vertex covering x (and as many other uncovered as possible)

    return D


procedure ADD_TO_D(v):
    if inD[v] == true: return
    inD[v] := true
    D.add(v)
    mark v and all neighbors of v as dominated
\end{verbatim}

The key design philosophy of \texttt{hedom5} is to combine:
\begin{itemize}
    \item \textbf{Very fast constructive heuristics} (reductions + lazy greedy),
    \item \textbf{Deterministic structural clean-up} (reverse pruning),
    \item \textbf{Targeted local improvement} (budgeted 1-swaps).
\end{itemize}
This is inspired by iterated-greedy / destroy-and-repair strategies in recent work on dominating set heuristics, which show that repeatedly rebuilding and locally repairing tends to outperform a single-shot greedy strategy on heterogeneous instances.

In practice, this pipeline consistently produced our best dominating sets under competition time limits, outperforming both the plain greedy baseline and greedy-seeded simulated annealing.

\bibliography{bibl}

\end{document}